\documentclass[11pt]{article}

\usepackage[preprint]{acl}

\usepackage{times}
\usepackage{latexsym}

\usepackage[T1]{fontenc}

\usepackage[utf8]{inputenc}

\usepackage{microtype}

\usepackage{inconsolata}

\usepackage{graphicx}

\usepackage{amsmath}
\usepackage{booktabs}
\usepackage{tabularx}
\usepackage{multirow}

\title{PhonoVoxtLM: Exploring Single-Codebook Phonological Tokens for Speech Understanding and Generation}

\author{
  Mingrui Liang \\
  Johns Hopkins University \\
  \texttt{mliang17@jh.edu}
  \And
  Kentaro Onda \\
  The University of Tokyo \\
  \And
  Shinji Watanabe \\
  Carnegie Mellon University \\
  \texttt{swatanab@andrew.cmu.edu}
}

\begin{document}
\maketitle

\begin{abstract}
We investigate how speech token representations can
support both understanding and generation within
a unified speech language model.
We present PhonoVoxtLM, built on the existing VoxtLM
framework using phonological tokens designed to retain
linguistic content and prosodic information.
The existing tokenizer is jointly trained for speech
recognition and resynthesis. Its single-codebook units serve as both speech inputs
and generation targets, without additional
speech-token prediction modules.
Experiments show improvements over HuBERT-based
baselines across understanding and generation under
different training-data scales.
Further comparisons with WavLM-based tokens reveal
trade-offs between recognition and synthesis,
highlighting the importance of evaluating shared
representations across both capabilities.
We further extend the framework to spoken question
answering through fine-tuning with replay.
The adapted model improves answering performance over
similarly adapted HuBERT baselines while retaining
advantages on the original speech tasks.
Our findings support single-codebook representations
designed for both content and prosody as a practical
choice for unified multitask speech modeling.
\end{abstract}

\section{Introduction}
\label{sec:introduction}

\begin{figure}[t]
    \centering
    \includegraphics[width=\columnwidth]{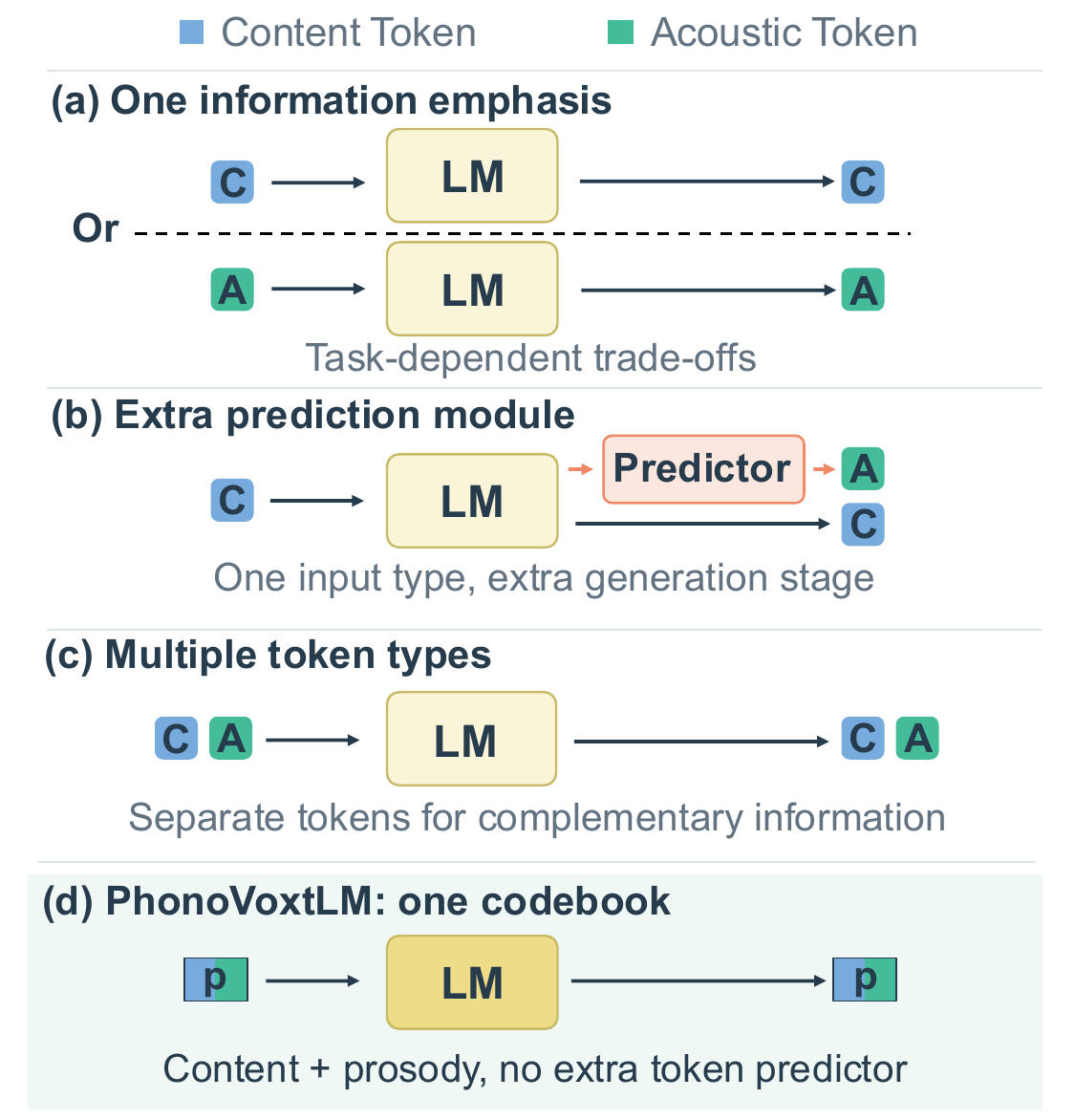}
    \caption{Speech-token modeling designs:
(a) content- or acoustic-oriented tokens;
(b) an extra token predictor;
(c) multiple token types; and
(d) phonological tokens designed to retain content
and prosody in a single codebook.
C/A denote content-oriented/acoustic tokens. Text paths and waveform decoders are omitted.}
    \label{fig:token_designs}
\end{figure}

Recent advances in text language models have established
autoregressive modeling as an effective framework for
diverse NLP tasks
\citep{radford2019language,brown2020language},
motivating efforts to unify speech understanding and
generation
\citep{zhang2023speechgpt,maiti2024voxtlm,nguyen2025spirit}.
Discrete speech tokens enable speech and text to be
modeled within a shared architecture.
When the same speech representation serves as both input
and generation target, it must accommodate the informational
demands of both capabilities
\citep{mousavi2025discrete,yang2024towards,zhang2024speechtokenizerunifiedspeechtokenizer}.

Existing systems address these demands through different
representation and modeling choices.
VALL-E uses acoustic codec tokens for synthesis
\citep{wang2023neural}, whereas GSLM models
content-oriented units for text-free speech generation
\citep{lakhotia2021generative} (Figure~\ref{fig:token_designs}(a)).
Such representations emphasize different aspects of
speech and may not serve understanding and generation
equally well.
Other systems introduce complementary mechanisms:
DualSpeechLM uses AcousticGPT to predict acoustic tokens
from understanding-oriented representations
\citep{wang2026dualspeechlm} (Figure~\ref{fig:token_designs}(b)), while SpiRit-LM Expressive
combines phonetic, pitch, and style tokens
\citep{nguyen2025spirit} (Figure~\ref{fig:token_designs}(c)).
These alternatives motivate our question:
how effectively can a single-codebook representation
support both understanding and generation within
a shared autoregressive framework?

We study this question with PhonoVoxtLM (Figure~\ref{fig:token_designs}(d)), built on
VoxtLM's unified framework for ASR, TTS, and speech
and text continuation \citep{maiti2024voxtlm}.
We replace its HuBERT units with tokens produced by
the existing Phonological Tokenizer \citep{onda2026phonological},
which combines ASR and speech resynthesis objectives
through differentiable $k$-means to retain linguistic
content and prosodic information.
The resulting units serve as both speech inputs and
generation targets, retaining VoxtLM's architecture
without additional token prediction modules.
Our comparisons examine codebook size, speech encoder
choice, and training-data scale to assess the benefits
and trade-offs of this representation.
We further extend the framework to spoken question
answering (SQA), testing question-conditioned
understanding beyond the original tasks.
We adapt the model with replay and evaluate both
SQA performance and original-task retention. To facilitate reproducibility, we implement our training and inference pipelines on ESPnet recipes \citep{watanabe2018espnet}.

Our core contributions are as follows:
\begin{itemize}
    \item We systematically evaluate phonological tokens
    as a shared input--output representation, finding
    improvements over 1K and 2K HuBERT baselines across
    understanding and generation under two data settings.

    \item We identify a trade-off relative to WavLM-large
$k$-means units: phonological tokens improve synthesis,
while the WavLM baselines achieve better recognition
and speech discrimination.

    \item We extend the shared interface to SQA and
    demonstrate gains over similarly adapted HuBERT
    baselines while retaining advantages on the
    original speech tasks.
\end{itemize}

\begin{figure*}[t]
    \centering
    \includegraphics[width=\textwidth]{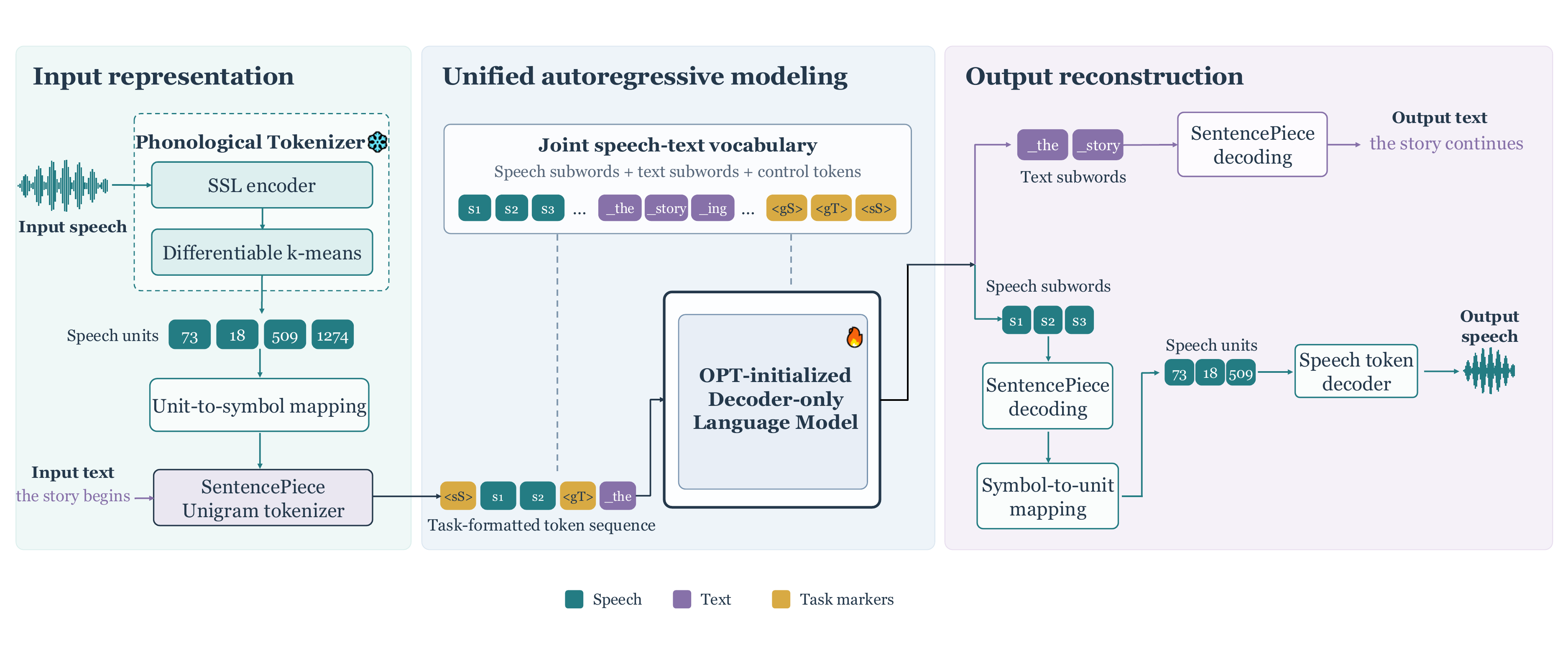}
    \caption{\textbf{Overview of PhonoVoxtLM within the VoxtLM framework.}
    A frozen Phonological Tokenizer extracts discrete
    speech units, which are mapped to symbols and
    processed by a SentencePiece unigram tokenizer
    shared with text.
    Task-formatted sequences are modeled by one
    OPT-based decoder-only language model.
    Generated text is detokenized directly; generated
    speech subwords are decoded and mapped back to
    speech units before waveform synthesis with HiFi-GAN.}
    \label{fig:architecture}
\end{figure*}

\section{Related Work}
\label{sec:related_work}

\subsection{Speech Tokenization}
\label{sec:rw_speech_tokenization}

Speech tokenizers convert continuous speech into discrete
units and are commonly categorized as acoustic or semantic
\citep{borsos2023audiolm,parker2025scaling,yang2024uniaudio}.
Acoustic tokenizers emphasize signal reconstruction,
preserving detailed acoustic information for generation
\citep{zeghidour2021soundstream,defossez2022high,yang2023hifi}.
However, these objectives do not necessarily yield
representations well suited to speech understanding
\citep{shi2024versaversatileevaluationtoolkit,chang2024exploring}.
In contrast, semantic tokenizers emphasize linguistic
content, typically by discretizing self-supervised
speech features \citep{hsu2021hubert,chen2022wavlm}
through clustering
\citep{borsos2023audiolm,zhang2023speechgpt,shi2024multi}
or learned vector quantization \citep{huang2024repcodec}.
Another approach introduces quantization bottlenecks
into ASR models \citep{du2024cosyvoice,zeng2024glm}. 

Recent work explores representations that accommodate
both linguistic and acoustic information, including
multi-codebook designs
\citep{zhang2024speechtokenizerunifiedspeechtokenizer,defossez2024moshi,yang2025almtokenizer}
and ReasoningCodec \citep{yang2026uniaudio}.
The Phonological Tokenizer \citep{onda2026phonological}
combines ASR and speech resynthesis objectives through
differentiable $k$-means and is designed to retain linguistic
content and prosodic information in a single codebook.
We adopt this existing phonological tokenizer and evaluate its units
as a shared input--output representation within VoxtLM,
focusing on their effects on both understanding and generation.

\subsection{Speech Language Models}
\label{sec:rw_speechtext_models}


Discrete-token speech language models formulate speech
modeling as autoregressive sequence prediction
\citep{borsos2023audiolm,wang2023neural,kharitonov2023speak,yang2023uniaudio}.
Several systems use representations with a particular
informational emphasis in task-focused settings.
VALL-E \citep{wang2023neural} models acoustic codec tokens
for text-to-speech synthesis, whereas GSLM
\citep{lakhotia2021generative} models discrete units derived
from self-supervised speech representations for text-free
speech generation.
These approaches demonstrate the effectiveness of discrete
speech modeling, but their representations are not necessarily
equally suited to the demands of both understanding
and generation.

To accommodate these differing demands, other systems
introduce specialized prediction components or combine
complementary token types.
DualSpeechLM \citep{wang2026dualspeechlm} uses
understanding-oriented tokens as language-model inputs
and introduces AcousticGPT to predict acoustic tokens
for generation.
Alternatively, SpiRit-LM Expressive \citep{nguyen2025spirit}
supplements phonetic units with pitch and style tokens,
while UniAudio 2.0 \citep{yang2026uniaudio} models reasoning
and reconstruction tokens for audio understanding
and generation. We build on VoxtLM's shared speech input--output framework
\citep{maiti2024voxtlm}.
We evaluate single-codebook phonological tokens within
this architecture, without introducing additional
speech-token prediction modules or complementary
speech-token streams.



\section{Method}
\label{sec:method}

We build on the unified autoregressive framework of VoxtLM \cite{maiti2024voxtlm} and investigate the use of phonological tokens as a
shared speech representation for model inputs and generation targets.
As illustrated in Figure~\ref{fig:architecture}, a frozen Phonological
Tokenizer converts speech into discrete units, which are incorporated into
a joint speech--text vocabulary through subword modeling. A single
OPT-initialized decoder-only language model predicts sequences over this
vocabulary. We retain the original text language modeling, speech language
modeling, automatic speech recognition (ASR), and text-to-speech synthesis
(TTS) tasks, and extend the framework to spoken question answering (SQA)
using the same sequence interface.

\subsection{Shared Phonological Speech Representation}
\label{sec:representation}

Given a waveform $X$, a speech tokenizer $\mathcal{T}_{\phi}$ produces a
sequence of discrete units:
\begin{equation}
    \begin{aligned}
        D &= \mathcal{T}_{\phi}(X) = (d_1,\ldots,d_N), \\
        d_i &\in \{0,\ldots,K-1\}.
    \end{aligned}
    \label{eq:speech_tokenization}
\end{equation}
We adopt the pretrained Phonological Tokenizer of
\citet{onda2026phonological}, which combines ASR and speech resynthesis
objectives through differentiable $k$-means to encourage the retention of
linguistic content and prosodic information. Our implementation uses its
fine-tuned WavLM-large encoder, extracting features from layer~21 and
quantizing them with the released $K=2{,}000$ cluster centroids. Both the
encoder and centroids remain frozen during language-model training and
SQA adaptation.

To integrate speech units with text, we follow VoxtLM and map each discrete speech-unit index $d_i$
to a unique character in the CJK Unified Ideographs block,
with Unicode code point $\texttt{0x4E00}+d_i$.
This one-to-one mapping $m$ is reversible and allows speech
units to be processed by SentencePiece alongside text. We then
train a single SentencePiece unigram tokenizer $U$
\citep{kudo2018sentencepiece} on sequences containing mapped speech units
and text from the ASR and TTS training data. This produces a joint
subword vocabulary containing speech symbols and their subsequences,
text subwords, and task-control tokens. SentencePiece is trained before
the language model and kept fixed thereafter. We denote the resulting
speech subword sequence by $S=U(m(D))$ and the tokenized text by $T$.
The 2,000-entry speech codebook is distinct from the joint subword
vocabulary: the former defines the shared discrete speech units, while
the latter defines the prediction space of the language model.

\subsection{Unified Autoregressive Multitask Modeling}
\label{sec:multitask}

We initialize the backbone of our decoder-only Transformer with
pretrained OPT weights \citep{zhang2022opt}, following VoxtLM,
which reports improved performance with pretrained
initialization compared to training from scratch
\citep{maiti2024voxtlm}. All tasks share the same language-model parameters and joint
output vocabulary. Accordingly, four control tokens specify the
conditioning and output modalities. We abbreviate the speech and text
condition markers as $\langle\mathrm{sS}\rangle$ and
$\langle\mathrm{sT}\rangle$, and the speech and text generation markers
as $\langle\mathrm{gS}\rangle$ and $\langle\mathrm{gT}\rangle$.
Table~\ref{tab:task_formats} summarizes the sequence formats.

\begin{table*}[t]
\centering
\small
\setlength{\tabcolsep}{4pt}
\renewcommand{\arraystretch}{1.25}

\newcommand{\ctok}[1]{%
    \ensuremath{\langle\texttt{#1}\rangle}%
}

\begin{tabularx}{\textwidth}{@{}l|X|X|c@{}}
    \toprule
    \multirow{2}{*}{Task}
    & \multicolumn{1}{c|}{\multirow{2}{*}{Training}}
    & \multicolumn{2}{c}{Inference} \\
    & & \multicolumn{1}{c}{Condition} & Prediction \\
    \midrule

    TextLM
    & $\ctok{generate-text},T$
    & $\ctok{generate-text},T_{\mathrm{pre}}$
    & $\hat{T}$ \\

    SpeechLM
    & $\ctok{generate-speech},S$
    & $\ctok{generate-speech},S_{\mathrm{pre}}$
    & $\hat{S}$ \\

    ASR
    & $\ctok{start-speech},S,\ctok{generate-text},T$
    & $\ctok{start-speech},S^{\mathrm{test}},
       \ctok{generate-text}$
    & $\hat{T}$ \\

    TTS
    & $\ctok{start-text},T,\ctok{generate-speech},S$
    & $\ctok{start-text},T^{\mathrm{test}},
       \ctok{generate-speech}$
    & $\hat{S}$ \\

    \addlinespace[4pt]

    SQA
    & \begin{tabular}[c]{@{}l@{}}
        $\ctok{start-speech},S,\ctok{start-text},$ \\
        $Q,\ctok{generate-text},A$
      \end{tabular}
    & \begin{tabular}[c]{@{}l@{}}
        $\ctok{start-speech},S^{\mathrm{test}},
          \ctok{start-text},$ \\
        $Q^{\mathrm{test}},\ctok{generate-text}$
      \end{tabular}
    & $\hat{A}$ \\

    \bottomrule
\end{tabularx}

\caption{\textbf{Sequence formats for training and inference.}
$S$ and $T$ denote speech and text subword sequences,
and $Q$ and $A$ denote questions and answers.
For TextLM and SpeechLM, inference conditions on an
observed prefix ($T_{\mathrm{pre}}$ or $S_{\mathrm{pre}}$)
and predicts its continuation.
Training loss is computed only after the generation marker.
End-of-sequence tokens are omitted.}
\label{tab:task_formats}

\end{table*}

Let $\mathcal{V}$ denote the joint speech--text vocabulary,
including task-control and end-of-sequence tokens.
For a task-specific prefix
$C=(c_1,\ldots,c_M)\in\mathcal{V}^{M}$
and target $Y=(y_1,\ldots,y_L)\in\mathcal{V}^{L}$,
where $y_L$ is the end-of-sequence token, the model
factorizes the conditional probability as
\begin{equation}
    p_{\theta}(Y\mid C)
    = \prod_{t=1}^{L} p_{\theta}(y_t\mid C,y_{<t}).
    \label{eq:autoregressive}
\end{equation}
Training uses teacher forcing and label-smoothed cross-entropy over
target positions. The unsmoothed, token-normalized objective for one
example is
\begin{equation}
    \mathcal{L}_{\mathrm{AR}}
    = -\frac{1}{L}\sum_{t=1}^{L}
    \log p_{\theta}(y_t\mid C,y_{<t}).
    \label{eq:ar_loss}
\end{equation}
The prefix, including its generation marker, provides context but does
not contribute to the loss. For TextLM and SpeechLM, the training prefix
contains only the corresponding generation marker. At inference time,
an observed text or speech prefix can additionally condition continuation.

\paragraph{Output reconstruction.}
For text outputs, SentencePiece decoding converts
predicted subwords back into text.
For speech outputs, it expands predicted speech
subwords into a symbol sequence, after which $m^{-1}$
recovers the underlying speech-unit sequence.
A unit-conditioned HiFi-GAN vocoder
\citep{kong2020hifi} synthesizes waveforms from these units.
We train a separate vocoder for each speech tokenizer
using the same LJSpeech \cite{ljspeech17} data and training configuration.
Each vocoder remains fixed across language-model
training settings and SQA adaptation.
Thus, speech inputs and generation targets share
the same codebook, without an additional language
model or prediction branch for converting between
speech-token types.

\subsection{Extension to Spoken Question Answering}
\label{sec:sqa_method}

We formulate SQA as $p_{\theta}(A\mid S,Q)$ using
the sequence format in Table~\ref{tab:task_formats},
with supervision restricted to the answer and
end-of-sequence tokens.
LibriSQA Parts~I and~II share this answer-generation
objective: multiple-choice examples use the correct
answer text as the target and omit candidate options
from the input.

We fine-tune the four-task checkpoint on SQA mixed
with replay examples from ASR, TTS, TextLM, and
SpeechLM, retaining their original sequence formats.
Only the shared language model is updated;
the speech tokenizer and SentencePiece model remain
fixed. No new task-control tokens or prediction
modules are introduced.

\section{Experimental Setup}
\label{sec:experimental_setup}

\subsection{Experimental Design}
\label{sec:experimental_design}

We organize our experiments around an OPT-350M model
trained on the balanced mixture $\mathcal{D}_{\mathrm{Bal}}$.
We examine data scaling by increasing unpaired speech
and text data to obtain $\mathcal{D}_{\mathrm{3M}}$,
while keeping the backbone and paired training data fixed.
We also extend the balanced-data model to SQA through
fine-tuning with replay of the original tasks, evaluating
both SQA performance and retention of the original capabilities.


\subsection{Data}
\label{sec:experimental_data}

We use
LibriLight \cite{kahn2020libri} for speech-only modeling, the LibriSpeech \cite{panayotov2015librispeech} language-modeling
text corpus for text-only modeling, the 960-hour LibriSpeech training
set for ASR, and the LibriTTS \cite{zen2019libritts} and VCTK \cite{yamagishi2019cstr} training data for TTS.
The balanced mixture $\mathcal{D}_{\mathrm{Bal}}$ is
constructed from 300K speech-only utterances sampled
from LibriLight, 300K text-only examples sampled from
the LibriSpeech language-modeling corpus, 281K paired examples from LibriSpeech for ASR, and
399K paired examples from LibriTTS and VCTK for TTS.
In $\mathcal{D}_{\mathrm{3M}}$, each unpaired component is increased to
3M examples, while the paired data and development and test splits
remain unchanged.

For SQA adaptation, we combine LibriSQA Parts~I and~II
\citep{zhao2024librisqa} using the training format described in
Section~\ref{sec:sqa_method}, reserving a common set of speakers
for development across both parts. The resulting SQA training
set contains approximately 204K examples. We mix these with
65K replay examples from the original tasks: 25K ASR, 25K
TextLM, 10K TTS, and 5K SpeechLM examples. This yields
approximately 269K training examples, with SQA and replay
contributing 5.7M and 5.1M supervised target tokens,
respectively, achieving an approximately balanced token
distribution.

\subsection{Models and Training}
\label{sec:experimental_models}

We compare our phonological-token models with
baselines using $k$-means units extracted from the sixth
layer of HuBERT-base. Our primary baseline follows the
1,000-unit codebook setting evaluated in VoxtLM. Since the Phonological Tokenizer
uses a 2,000-unit codebook, we additionally include
2,000-unit HuBERT baselines to assess whether the observed
gains can be explained by the larger codebook alone.
Both HuBERT baselines are evaluated under the balanced
and expanded-data settings and after SQA adaptation.
Corresponding systems use the same backbone size
and task-data configuration. Our comparisons
evaluate the complete speech tokenization pipelines,
which differ in their encoders and training objectives.

Furthermore, we include WavLM-large $k$-means baselines
with 1,000 and 2,000 units, extracting features from the
21st layer, as in the Phonological Tokenizer.
These experiment use OPT-350M and $\mathcal{D}_{\mathrm{Bal}}$
with the same language-model training settings.

All models use a 10K SentencePiece unigram vocabulary
and an OPT-350M-initialized backbone. For the original four-task training,
we use Adam with a peak learning rate of $3\times10^{-4}$, 2,500 warmup
updates, and weight decay of $10^{-6}$. Training uses mixed precision,
gradient accumulation over 20 minibatches, gradient clipping at 1.0,
and label smoothing of 0.1. Checkpoints are selected by development-set
target-token accuracy.

For SQA adaptation, we reduce the peak learning rate to $10^{-4}$ and
use 300 warmup updates. We select the checkpoint with the highest
development-set answer-token accuracy and apply early stopping based
on development loss with a patience of four validation epochs.
The speech tokenizer and SentencePiece model remain fixed during
adaptation.

\begin{table*}[t]
\centering
\scriptsize
\setlength{\tabcolsep}{2pt}
\renewcommand{\arraystretch}{1.15}

\begin{tabular*}{\textwidth}{
@{\extracolsep{\fill}}llccccccccccc@{}}
\toprule
Setting & Speech tokens
& \multicolumn{3}{c}{TextLM}
& \multicolumn{3}{c}{SpeechLM}
& ASR
& \multicolumn{4}{c}{TTS} \\
\cmidrule(lr){3-5}
\cmidrule(lr){6-8}
\cmidrule(lr){9-9}
\cmidrule(lr){10-13}
& & PPL$\downarrow$ & sWUGGY$\uparrow$ & sBLIMP$\uparrow$
& PPL$\downarrow$ & sWUGGY$\uparrow$ & sBLIMP$\uparrow$
& WER$\downarrow$ & CER$\downarrow$ & UTMOS$\uparrow$
& SBS$\uparrow$ & F0 corr$\uparrow$ \\
\midrule

$\mathcal{D}_{\mathrm{Bal}}$
& HuBERT-1K
& 27.26 & \textbf{82.61} & 68.41
& 42.67 & 61.43 & 54.35
& 7.1/17.8 & 2.51 & 3.87 & 0.813 & 0.167 \\
& HuBERT-2K
& 23.78 & 82.52 & \textbf{68.97}
& 31.02 & 59.79 & 54.60
& 7.4/17.5 & 1.94 & 3.87 & 0.812 & 0.227 \\
& Phonological
& 26.17 & 82.16 & 68.83
& 36.76 & \textbf{66.98} & \textbf{57.31}
& \textbf{4.2}/\textbf{8.6}
& \textbf{1.61} & \textbf{3.98}
& \textbf{0.821} & \textbf{0.282} \\
\midrule

$\mathcal{D}_{\mathrm{Bal}}$
& HuBERT-1K
& 30.28 & \textbf{82.87} & 67.86
& 49.70 & 61.25 & 54.51
& 9.0/21.2 & 1.79 & 3.86 & 0.811 & 0.221 \\
+ SQA adaptation
& HuBERT-2K
& 26.70 & 82.27 & \textbf{68.92}
& 34.30 & 59.29 & 54.47
& 8.5/20.2 & 3.14 & 3.83 & 0.807 & 0.210 \\
& Phonological
& 29.90 & 82.23 & 68.86
& 43.20 & \textbf{66.39} & \textbf{57.03}
& \textbf{5.4}/\textbf{9.6}
& \textbf{1.50} & \textbf{3.93}
& \textbf{0.815} & \textbf{0.275} \\
\midrule

$\mathcal{D}_{\mathrm{3M}}$
& HuBERT-1K
& 19.52 & 84.22 & 74.08
& 34.06 & 64.78 & 56.25
& 6.9/16.6 & 2.46 & 3.80 & 0.807 & 0.199 \\
& HuBERT-2K
& 17.98 & 84.16 & 74.32
& 25.32 & 63.42 & 55.62
& 7.1/17.7 & 2.35 & 3.78 & 0.803 & 0.186 \\
& Phonological
& 19.00 & \textbf{84.43} & \textbf{74.63}
& 29.60 & \textbf{69.12} & \textbf{58.81}
& \textbf{4.3}/\textbf{8.3}
& \textbf{2.26} & \textbf{4.01}
& \textbf{0.813} & \textbf{0.278} \\
\bottomrule
\end{tabular*}

\caption{\textbf{Original-task results using OPT-350M.}
The first two blocks compare balanced-data models before
and after SQA fine-tuning with replay; the third uses
expanded unpaired training data.
HuBERT-1K and HuBERT-2K use 1,000- and 2,000-unit
codebooks, respectively; Phonological uses 2,000 units.
ASR WER is reported as test-clean/test-other.
Discrimination scores and error rates are percentages.
SBS denotes SpeechBERTScore; F0 corr denotes mean
F0 correlation.
Arrows indicate the preferred direction.
Within each block, the best values are bold,
except for PPL.}
\label{tab:main_results}
\label{tab:retention_results}
\end{table*}

\subsection{Evaluation}
\label{sec:experimental_evaluation}

\paragraph{Language modeling and recognition.}
We evaluate TextLM and SpeechLM using perplexity (PPL), and lexical
and syntactic discrimination using the sWUGGY and sBLIMP development
sets \citep{dunbar2022self}. The discrimination pairs are ranked by summed token
log-likelihood, without length normalization. SpeechLM is evaluated on speech
inputs, while TextLM uses the corresponding text. The discrimination
scores measure the percentage of pairs for which the model prefers
the valid word or grammatically acceptable sentence. Because the joint SentencePiece tokenizer is trained
separately for each speech tokenization system,
token-level PPL is not directly comparable across systems. For ASR, we report
word error rate (WER) separately on LibriSpeech test-clean and test-other.

\paragraph{Speech synthesis.}
We evaluate TTS on a fixed random subset of 100 test utterances.
With the help of VERSA \citep{shi2025versa}, we report character error rate (CER) from automatic transcription
of synthesized speech using Whisper large-v3 \citep{radford2022whisper}, UTMOS for predicted perceptual quality
\citep{saeki2022utmos}, SpeechBERTScore for similarity to reference
speech \citep{saeki2024speechbertscore}, and mean F0 correlation between synthesized and reference speech. Because synthesized and reference speech differ in
speaker identity, we interpret F0 correlation as a
comparative measure under matched evaluation conditions,
rather than a direct measure of speaker-matched
prosody preservation.
Decoding starts with beam search of width five and
no-repeat 8-gram blocking. Outputs that reach the
generation-length limit are decoded again using greedy
search, followed, if necessary, by nucleus sampling
with $p=0.9$ and temperature 1.0.

\paragraph{Spoken question answering.}
For LibriSQA Part~I, we evaluate free-form answer generation on a
fixed random subset of 500 test examples using BLEU-4, ROUGE-L F1,
and BERTScore F1 with baseline rescaling
\citep{papineni2002bleu,lin2004rouge,zhang2019bertscore}.
We use a decoding length penalty of 0.3, selected on development data.
We also report a transcript-as-answer baseline, which returns the
reference transcript without answering the question.

For Part~II, we report multiple-choice accuracy by scoring each
candidate answer as a continuation of the speech-and-question prefix.
Our main score subtracts the answer log-likelihood obtained without speech:
\begin{equation}
    \begin{aligned}
    s(A_j;S,Q) ={}& \log p_{\theta}(A_j\mid S,Q) \\
                 &-\log p_{\theta}(A_j\mid \emptyset,Q).
    \end{aligned}
    \label{eq:sqa_calibrated_score}
\end{equation}
Here, $\emptyset$ denotes an empty speech segment with the other
prompt elements retained. We select the candidate with the highest
score. Candidate log-likelihoods sum over answer tokens, excluding
the end-of-sequence token.
Finally, we re-evaluate the original four tasks after SQA adaptation
to assess changes in their performance.

\section{Results and Analysis}
\label{sec:results}








\begin{table*}[t]
\centering
\scriptsize
\setlength{\tabcolsep}{2pt}
\renewcommand{\arraystretch}{1.15}

\begin{tabular*}{\textwidth}{
@{\extracolsep{\fill}}lccccccccccc@{}}
\toprule
Speech tokens
& \multicolumn{3}{c}{TextLM}
& \multicolumn{3}{c}{SpeechLM}
& ASR
& \multicolumn{4}{c}{TTS} \\
\cmidrule(lr){2-4}
\cmidrule(lr){5-7}
\cmidrule(lr){8-8}
\cmidrule(lr){9-12}
& PPL$\downarrow$ & sWUGGY$\uparrow$ & sBLIMP$\uparrow$
& PPL$\downarrow$ & sWUGGY$\uparrow$ & sBLIMP$\uparrow$
& WER$\downarrow$ & CER$\downarrow$ & UTMOS$\uparrow$
& SBS$\uparrow$ & F0 corr$\uparrow$ \\
\midrule

WavLM-1K
& 28.66 & \textbf{82.34} & 68.92
& 33.85 & 68.12 & 58.07
& \textbf{3.7/6.5}
& 1.63 & 3.70 & 0.806 & 0.144 \\

WavLM-2K
& 26.79 & 81.73 & \textbf{69.51}
& 22.01 & \textbf{68.28} & \textbf{58.30}
& 3.8/6.8
& 3.03 & 3.62 & 0.795 & 0.159 \\

Phonological
& 26.17 & 82.16 & 68.83
& 36.76 & 66.98 & 57.31
& 4.2/8.6
& \textbf{1.61} & \textbf{3.98}
& \textbf{0.821} & \textbf{0.282} \\

\bottomrule
\end{tabular*}

\caption{\textbf{Comparison with WavLM-based $k$-means units
using OPT-350M and $\mathcal{D}_{\mathrm{Bal}}$.}
WavLM-1K and WavLM-2K use 1,000-unit and 2,000-unit
codebooks; Phonological uses 2,000 units.
WER is reported as test-clean/test-other.
Discrimination scores and error rates are percentages.
SBS denotes SpeechBERTScore; F0 corr denotes mean
F0 correlation.
Arrows indicate the preferred direction.
Best values are bold except for PPL.}
\label{tab:ablation_results}
\end{table*}

\subsection{Performance on the Original Four Tasks}
\label{sec:results_base}

Table~\ref{tab:main_results} shows that phonological
tokens improve both speech understanding and generation
over the HuBERT baselines within the same VoxtLM framework.
In the balanced-data setting, gains span speech
discrimination, recognition, and all four TTS metrics.
For example, ASR WER decreases from 7.1\%/17.8\%
with HuBERT-1K to 4.2\%/8.6\%, and F0 correlation
increases from 0.167 to 0.282.
This pattern supports the value of improving the shared
speech representation without adding speech-token
prediction modules, although it does not isolate
the contribution of prosodic information.



\paragraph{Controlling for codebook size.}
Increasing the HuBERT codebook from 1K to 2K improves
TTS CER but lowers speech sWUGGY under both data settings.
Phonological tokens outperform the size-matched
HuBERT-2K baseline in recognition, speech discrimination,
and synthesis, indicating that codebook size alone
does not explain the joint improvements.

\paragraph{Effects on text modeling.}
Text discrimination remains broadly comparable,
suggesting that the speech gains do not come with
a consistent loss in text discrimination.
We report PPL for completeness but do not compare it
across tokenizers because their tokenizations differ.


\subsection{Data Scaling}
\label{sec:results_scaling}

\begin{table*}[t]
\centering
\scriptsize
\setlength{\tabcolsep}{4pt}
\renewcommand{\arraystretch}{1.2}

\begin{tabular*}{\textwidth}{@{\extracolsep{\fill}}llcccc@{}}
\toprule
& & \multicolumn{3}{c}{Part I: answer generation} & Part II \\
\cmidrule(lr){3-5}
\cmidrule(lr){6-6}
Speech tokens & SQA fine-tuning
& BLEU-4$\uparrow$ & ROUGE-L F1$\uparrow$
& BERTScore F1$\uparrow$ & Accuracy$\uparrow$ \\
\midrule
HuBERT-1K & No
& 21.30 & 41.10 & 46.50 & \textbf{36.80} \\
HuBERT-2K & No
& 21.90 & 41.60 & 47.40 & 34.80 \\
Phonological & No
& \textbf{22.82} & \textbf{42.28} & \textbf{49.14} & 35.73 \\
\midrule
HuBERT-1K & Yes
& 29.50 & 49.00 & 55.40 & 45.10 \\
HuBERT-2K & Yes
& 27.78 & 47.48 & 54.08 & 44.77 \\
Phonological & Yes
& \textbf{30.75} & \textbf{49.01}
& \textbf{55.64} & \textbf{46.76} \\
\midrule
\multicolumn{2}{l}{Transcript-as-answer baseline}
& 23.71 & 43.54 & 51.29 & --- \\
\bottomrule
\end{tabular*}

\caption{\textbf{LibriSQA results from OPT-350M models trained on balanced
data.} Part I uses the fixed 500-example test subset. BLEU-4,
ROUGE-L F1, and rescaled BERTScore F1 are multiplied by 100 for reporting.
Part II accuracy (\%) uses the No-Audio-Calibrated Answer Score
for candidate selection. Best values are bold separately within
the three models before and after SQA fine-tuning.}
\label{tab:sqa_results}
\end{table*}

The advantages of phonological tokens over both HuBERT
baselines persist under $\mathcal{D}_{\mathrm{3M}}$
(Table~\ref{tab:main_results}).
Gains remain consistent across recognition, speech
discrimination, and all four TTS metrics, suggesting
that the benefits of the shared representation are
not confined to the balanced-data setting.

However, scaling unpaired data yields uneven gains
across tasks.
For the phonological model, text and speech discrimination
improve, whereas recognition changes little and synthesis
metrics show mixed effects.
In particular, TTS CER increases from 1.61\% to 2.26\%,
despite a small improvement in UTMOS.
Thus, the representation retains its relative advantages,
but additional unpaired data do not uniformly improve
understanding and generation.

\subsection{Understanding--Generation Trade-offs}
\label{sec:results_ablation}



Table~\ref{tab:ablation_results} reveals a trade-off
between understanding and generation when phonological
tokens are compared with WavLM-large $k$-means units
extracted from the same layer.
The WavLM baselines achieve higher speech discrimination
scores and lower ASR WER, while phonological tokens
perform better on all four TTS metrics.

The synthesis advantage is most evident in predicted
quality and reference pitch correlation.
Phonological tokens achieve a UTMOS of 3.98 and F0
correlation of 0.282, compared with 3.62--3.70 and
0.144--0.159 for the WavLM baselines.
The CER of PhnonVoxtLM is close to WavLM-1K and lower than WavLM-2K.
These results highlight why a shared representation
should be evaluated across both capabilities:
the units with stronger recognition and discrimination
scores do not necessarily provide stronger synthesis.
Phonological tokens offer a useful balance, improving
synthesis while retaining competitive understanding
performance.

\subsection{Spoken Question Answering}
\label{sec:results_sqa}

SQA adaptation extends the shared token interface to
question-conditioned answer generation without additional
prediction modules, using the sequence format in
Table~\ref{tab:task_formats}.
Table~\ref{tab:sqa_results} shows improvements in both
answer generation and selection; for the phonological
model, Part~II accuracy rises from 35.73\% to 46.76\%.

Among similarly adapted models, phonological tokens
achieve higher BLEU-4 and BERTScore F1 than HuBERT-1K,
with nearly identical ROUGE-L F1, and exceed HuBERT-2K
on all three generation metrics.
Using the calibrated score in
Equation~\ref{eq:sqa_calibrated_score}, Part~II accuracy
is higher by 1.66 and 1.99 percentage points,
respectively.
The adapted model also exceeds the transcript-as-answer
baseline on all Part~I metrics, producing answers
better matched to the references than simply returning
the transcript.
These results extend the evidence for the shared
representation beyond recognition and discrimination
to spoken question answering.

\subsection{Retention of the Original Tasks}
\label{sec:results_retention}



SQA adaptation with replay affects the original tasks
unevenly (the first two blocks of
Table~\ref{tab:main_results}).
For the phonological model, recognition and speech
discrimination decline, and TextLM PPL increases.
For example, ASR WER rises from 4.2\%/8.6\% to
5.4\%/9.6\%.
TTS CER improves, while UTMOS, SpeechBERTScore, and
F0 correlation decrease slightly.

Nevertheless, the adapted model remains ahead of both
adapted HuBERT baselines on recognition, speech
discrimination, and all four TTS metrics.
The representation therefore retains its comparative
advantages after adaptation, although the model does
not fully preserve its pre-adaptation performance.

\begin{table}[t]
\centering
\scriptsize
\setlength{\tabcolsep}{2pt}
\renewcommand{\arraystretch}{1.1}

\begin{tabular*}{\columnwidth}{
@{\extracolsep{\fill}}lcccc@{}}
\toprule
Model & Params
& \multicolumn{2}{c}{SpeechLM}
& ASR \\
\cmidrule(lr){3-4}
& & sWUGGY$\uparrow$ & sBLIMP$\uparrow$
& WER$\downarrow$ \\
\midrule
GSLM-$k200$
& 172M & --- & 53.0 & --- \\
AudioLM
& 900M & 71.5 & 64.7 & --- \\
VoxtLM-$k200$
& 1.3B & 66.1 & 56.7 & 4.6/12.1 \\
\quad + MLS
& 1.3B & 65.6 & 57.1 & 2.7/6.5 \\
TWIST
& 1.3B & 72.7 & 57.0 & --- \\
SpiRit-LM Base 
& 7B & 71.4 & 63.2 & 21.9/29.2 \\
\quad + ASR + TTS
& 7B & 71.8 & 62.4 & 6.0/11.0 \\
\midrule
\textbf{PhonoVoxtLM}$^{a}$
& 350M & 69.1 & 58.8 & 4.3/8.3 \\
\textbf{PhonoVoxtLM}$^{b}$
& 350M & 66.4 & 57.0 & 5.4/9.6 \\
\bottomrule
\end{tabular*}

\caption{\textbf{Comparison of our models against prior work.}
GSLM, AudioLM, VoxtLM, TWIST and SpiRit-LM results are reported in \citet{lakhotia2021generative}, \citet{borsos2023audiolm},
\citet{maiti2024voxtlm}, \citet{hassid2023textually} and
\citet{nguyen2025spirit}, respectively. SpiRit-LM ASR uses 10-shot prompting.
ASR WER is reported as LibriSpeech test-clean/test-other.
Training data and evaluation protocols differ across studies.
$^{a}$Trained on $\mathcal{D}_{\mathrm{3M}}$.
$^{b}$Trained on $\mathcal{D}_{\mathrm{Bal}}$,
then fine-tuned on SQA with replay.}
\label{tab:published_comparison}
\end{table}

\subsection{Comparison with Prior Work}
\label{sec:results_prior_work}

Table~\ref{tab:published_comparison} places PhonoVoxtLM
in the context of prior systems, although training data
and evaluation protocols differ.
Our 350M model trained on $\mathcal{D}_{\mathrm{3M}}$
achieves higher reported speech discrimination scores
and lower ASR WER than the 1.3B VoxtLM-$k200$
without additional MLS data.
Although other systems lead on individual metrics,
PhonoVoxtLM achieves a competitive balance between
speech modeling and recognition with a relatively
small language-model backbone.

\section{Conclusion}
\label{sec:conclusion}

We investigated phonological tokens as a shared
input--output representation within the existing
VoxtLM framework.
Across balanced and expanded-data settings, these
tokens improve speech recognition, discrimination,
and synthesis over HuBERT baselines.
Comparisons with WavLM-based units reveal a trade-off:
phonological tokens provide stronger synthesis,
while WavLM units perform better in recognition
and speech discrimination.
After SQA adaptation, the phonological model also
outperforms similarly adapted HuBERT baselines
while retaining advantages on the original speech tasks.
Together, these findings support single-codebook
phonological tokens as a practical shared representation
for understanding and generation, without additional
speech-token prediction modules.

\section*{Limitations}

Our experiments compare complete speech representation
pipelines. The WavLM baselines match the encoder
architecture and feature-extraction layer, but do not
isolate individual tokenizer training objectives.
F0 correlation provides evidence about reference pitch
matching, but does not establish the causal contribution
of prosodic information to downstream performance
or assess prosody comprehensively.
Our evaluation
is also limited to English, primarily read speech, with
spoken question answering assessed only on LibriSQA.
Further studies are needed to examine generalization to
other languages and conversational settings. Besides, our experiments use a single language-model size;
generalization to larger backbones remains to be evaluated. Finally,
speech synthesis is evaluated using automatic metrics;
human listening tests would provide complementary evidence
on perceived naturalness and prosodic quality.

\section*{Acknowledgments}

We thank the Center for Language and Speech Processing
(CLSP) at Johns Hopkins University for providing the
computational resources used in our experiments.


\bibliography{custom}




\end{document}